\documentclass[conference]{IEEEtran}
\IEEEoverridecommandlockouts
\usepackage{cite}
\usepackage{amsmath,amssymb,amsfonts}
\usepackage{graphicx}
\usepackage{textcomp}
\usepackage{algpseudocode}
\usepackage{algorithm}
\usepackage{booktabs}
\usepackage{listings}
\usepackage{csquotes}
\usepackage{hyperref}
\usepackage{xcolor}
\usepackage{fontawesome}
\usepackage{cleveref}
\usepackage{array}

\crefformat{figure}{Fig.~#2#1#3}

\def\BibTeX{{\rm B\kern-.05em{\sc i\kern-.025em b}\kern-.08em
    T\kern-.1667em\lower.7ex\hbox{E}\kern-.125emX}}

\begin{document}

\title{
A 3D Framework for Improving Low-Latency \\
Multi-Channel Live Streaming
}

\author{\IEEEauthorblockN{
Aizierjiang Aiersilan
}
\IEEEauthorblockA{
University of Macau\\
Macau SAR, China\\
ezharjan@outlook.com
}
\and
\IEEEauthorblockN{
Zhiqiang Wang
}
\IEEEauthorblockA{
Beijing University of Technology\\
Beijing, China\\
wzqcg@bjut.edu.cn}
}

\maketitle

\begin{abstract}
The advent of 5G has driven the demand for high-quality, low-latency live streaming. However, challenges such as managing the increased data volume, ensuring synchronization across multiple streams, and maintaining consistent quality under varying network conditions persist, particularly in real-time video streaming. To address these issues, we propose a novel framework that leverages 3D virtual environments within game engines (e.g., Unity 3D) to optimize multi-channel live streaming. Our approach consolidates multi-camera video data into a single stream using multiple virtual 3D canvases, significantly increasing channel amounts while reducing latency and enhancing user flexibility. For demonstration of our approach, we utilize the Unity 3D engine to integrate multiple video inputs into a single-channel stream, supporting one-to-many broadcasting, one-to-one video calling, and real-time control of video channels. By mapping video data onto a world-space canvas and capturing it via an in-world camera, we minimize redundant data transmission, achieving efficient, low-latency streaming. Our results demonstrate that this method outperforms some existing multi-channel live streaming solutions in both latency reduction and user interaction responsiveness improvement. 
\end{abstract}


\section{Introduction}
\label{sec:intro}
\setlength{\parindent}{0pt}
Live video streaming is crucial in digital communication, encompassing social media, virtual events, and online education. The advent of 5G networks has heightened the demand for high-quality, low-latency streaming by reducing delays and enhancing data transmission. However, challenges remain in managing large data volumes, synchronizing multiple video streams, and maintaining quality under fluctuating network conditions—issues particularly critical in real-time streaming where even slight delays impair user experience and system performance.

To address these issues, Chen et al. \cite{chen2024videollm} proposed a framework offering solutions for video streaming dialogue. However, implementing 3D virtual environments in live streaming introduces significant obstacles. Real-time processing and rendering of multiple video streams demand substantial computational power, especially in complex virtual settings. Maintaining synchronization across video sources is technically challenging, as consistent timing is vital to avoid desynchronization. Our approach ensures synchronization by rendering all video streams within a unified 3D scene in Unity 3D (U3D), ensuring consistent frame timing during capture, as detailed in \cref{sec:sync_accuracy}. Furthermore, managing low latency with high data volumes from multi-channel setups remains difficult under variable network conditions.

Our approach utilizes 3D virtual environments within engines like U3D to optimize multi-channel live streaming. This method integrates video data from multiple cameras within a virtual 3D space, mapping each stream onto a virtual canvas — a 3D surface displaying video content — and capturing it with an in-world camera. By consolidating multi-camera data into a single stream, our approach reduces complexity and minimizes redundant data transmission. Leveraging U3D for real-time rendering and data management, our method decreases latency and enhances flexibility in video delivery.

Using 3D virtual environments for live streaming offers capabilities beyond traditional methods. Unlike conventional techniques, our approach enables a dynamic, interactive streaming experience where users can select different camera inputs, angles, or perspectives in real-time, creating a more immersive experience. This method supports the demand for interactive content in virtual reality (VR), augmented reality (AR), and mixed reality (MR), where real-time processing and low latency are critical. Consequently, our approach extends its impact to remote collaboration, virtual education, and telemedicine.

While existing methods may reduce latency, they often require significant bandwidth, which can be a bottleneck in less robust network environments. Our framework optimizes video data processing within virtual 3D environments to enhance scalability and reduce computational load. By efficiently mapping video data onto virtual canvases and using in-world cameras to consolidate streams, our method minimizes redundant processing and reduces data volume transmitted across the network. This approach not only reduces latency but also offers a more flexible and scalable solution for multi-channel live streaming. Experimental validation shows our method outperforms existing solutions in latency, user interaction, and system scalability.

We make 4 major contributions:

\textbullet\hspace{0.5em} Latency reduction of up to 68.7\% compared to existing multi-channel live streaming methods.

\textbullet\hspace{0.5em} Improved scalability by optimizing multi-camera video data processing and transmission.

\textbullet\hspace{0.5em} Enhanced user interaction through real-time control within a 3D virtual environment, with response times averaging 600ms.

\textbullet\hspace{0.5em} Broader applicability across VR/AR/MR and remote collaboration, providing a robust solution for diverse industry needs.

\section{Related Work}
The rapid expansion of live video streaming has driven extensive research to tackle challenges in multi-channel streaming, particularly regarding latency, synchronization, scalability, and video quality.

\subsection{Latency Reduction in Live Streaming}
Latency is critical in live streaming, especially for real-time interactions in gaming, telemedicine, and VR/AR/MR. Strategies to minimize latency include integrating Model Predictive Control with deep reinforcement learning, achieving 2 to 5 seconds of latency \cite{sun2021towards}. However, this is high for real-time applications. The \textit{Learn2Adapt-LowLatency} achieves 1.04 seconds latency without parameter tuning \cite{karagkioules2020online}, but is insufficient for multi-channel scenarios. Adaptive Bitrate Streaming (ABR) dynamically adjusts video quality to match bandwidth, reducing buffering delays but introducing latency due to quality adjustments \cite{tashtarian2024artemis, bin2021new, turkkan2022greenabr, duanmu2020assessing}. \textit{MultiLive} reduces end-to-end delay to 100 ms \cite{wang2021multilive}, but bandwidth increases with channel count. Recent approaches like LLL-CAdViSE \cite{taraghi2023lll}, FastEmit \cite{yu2021fastemit}, HxL3 \cite{tashtarian2022mathsf}, and LiveNet \cite{li2022livenet} focus on ultra-low-latency but require significant resources and are less suitable for multi-channel scenarios. Our method uses U3D to consolidate multiple streams, reducing latency while maintaining synchronization.

\subsection{Synchronization Across Multiple Streams}
Synchronization across multiple video streams is a major challenge, as minor discrepancies cause desynchronization artifacts. Traditional methods use timestamps for alignment, but varying network delays pose issues. \textit{MMT-based Multi-channel Video Transmission System} \cite{mochida2020mmt} and cloud-edge collaboration \cite{guo2022multi} enhance synchronization but increase computational complexity and latency. Our framework centralizes synchronization within the 3D virtual environment, ensuring consistent timing with minimal overhead, as detailed in \cref{sec:architecture}.

\subsection{Scalability in Multi-Channel Streaming}
Scalability is crucial as the number of video streams grows. Traditional architectures struggle to scale efficiently, as each stream adds to computational and bandwidth demands. Scalable Video Coding \cite{Shahid11} allows incremental decoding for better bandwidth use but introduces latency and complexity. Edge computing (\cite{ali2020res, ma2022qava, gao2024low}, etc.) offloads processing closer to data sources, reducing server load and improving responsiveness, but requires substantial infrastructure. Our approach optimizes video data processing within a virtual 3D environment, efficiently managing multiple streams without extensive infrastructure.

\subsection{User Interaction in Real-Time Streaming}
User interaction is increasingly important in immersive environments like VR and AR, where real-time control and feedback are crucial. Traditional architectures struggle to provide responsiveness, leading to disruptive delays. Some approaches \cite{nacakli2020controlling, farahani2022ararat, shen20215g} reduce these delays by streaming from edge servers, but require significant infrastructure and may not scale across regions. Our framework enhances interaction by integrating real-time control within the 3D virtual environment, allowing immediate updates based on user inputs without extensive infrastructure.

\subsection{Advantages of Our Approach}
Despite advancements in multi-channel live streaming, existing methods often compromise on latency, synchronization, scalability, and video quality. Our framework harnesses 3D game engines to consolidate video streams, reducing latency, improving synchronization, enhancing scalability, and maintaining an acceptable quality under varying conditions while enabling real-time user interaction.

\section{3D Framework for Live Video Streaming}
Our method adopts the strategy of multi-channel/multi-input data fusion in 3D space, leveraging the U3D engine to map video streams onto virtual canvases, which are then captured by an in-world camera to create a consolidated data stream. To aid clarity, we define U3D-specific terms: a \textit{virtual canvas} is a 3D surface in the virtual environment displaying video content; a \textit{Raw Image display board} is an object showing a single video stream; an \textit{interactive control group} is a set of interactive elements controlling stream display or user interactions. Our framework addresses latency reduction, synchronization, and scalability in real-time video streaming.

\subsection{Design Principles}
The design of our framework is grounded in three core principles: modularity and scalability, low latency, and spatial awareness.

\textbf{Modularity and Scalability.} 
Our framework is structured with a modular architecture, where each video channel operates independently. Such modularity ensures the system's scalability, allowing it to handle an increasing number of video channels without significant performance degradation. The modular approach also enhances flexibility, enabling live streaming systems to adapt to various operational environments and requirements.

\textbf{Low Latency.} 
Latency is minimized in multi-channel scenarios using our method, which ensures that the communication latency for multi-channel video streaming is equivalent to that of single-channel streaming. This reduction in latency is achieved by streaming multiple input video streams through a single channel, effectively lowering the communication latency typically associated with multi-channel streaming.

\textbf{Spatial Awareness.} 
By employing a 3D virtual environment, our framework enhances the processing and display of multiple video streams. Video data is mapped onto a virtual canvas within the 3D space, and different virtual cameras' capabilities inside U3D are utilized for real-time rendering and manipulation. This spatial approach not only improves user experience but also facilitates more intuitive interactions with multi-channel content.

\begin{figure}[ht]
    \centering
    \vspace{-4mm}
    \includegraphics[width=\linewidth]{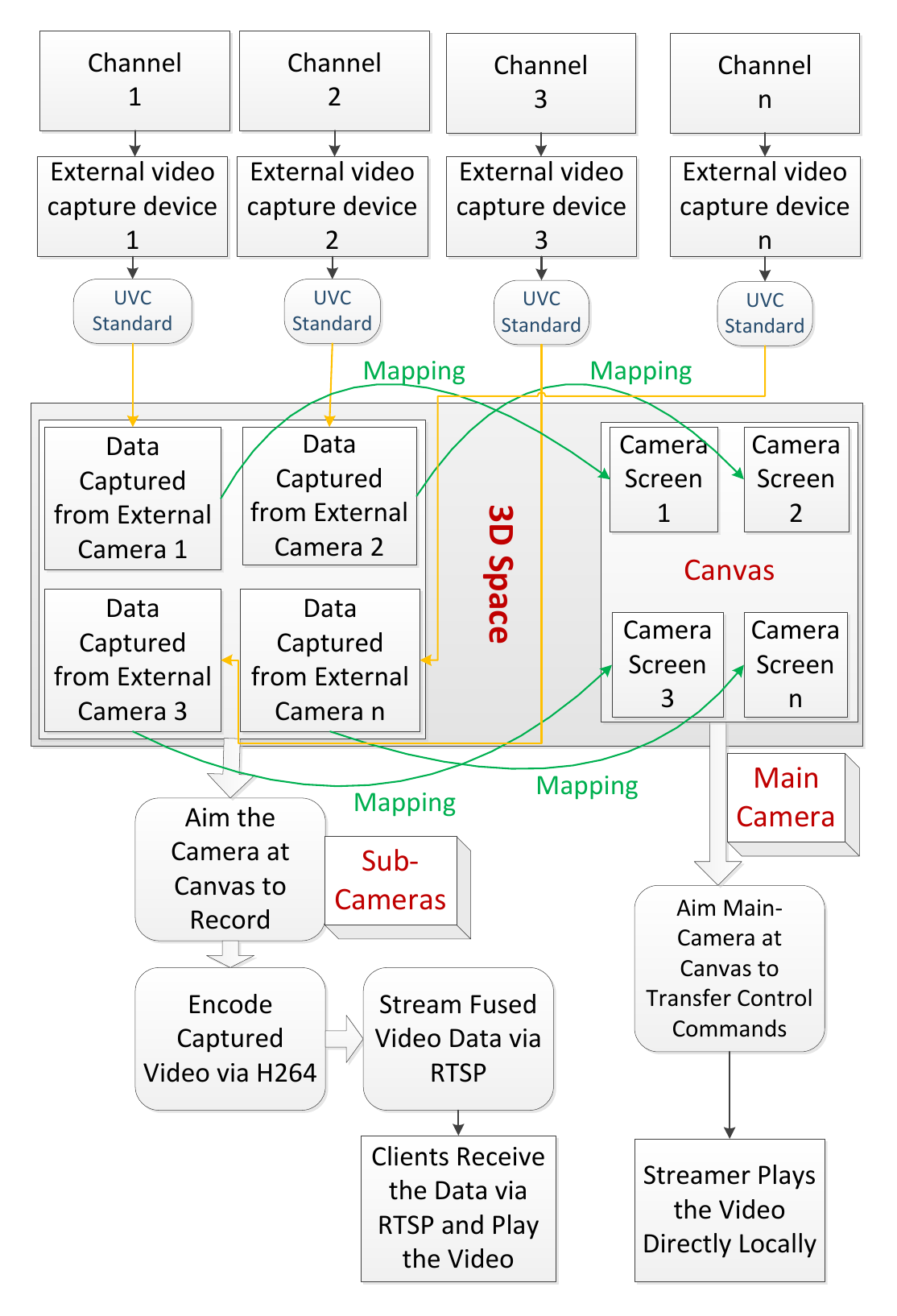}
    \vspace{-1cm}
    \caption{Our framework leverages multiple input devices to create channels for clients. Video content is seamlessly distributed across canvases in a virtual 3D environment, captured by sub-cameras and then consolidated by a main camera. Users' interactions with the cameras are mapped to an interactive control group, ensuring a responsive experience. The final stream is transmitted to clients via RTSP. To acquire video data from external cameras, we utilize the UVC standard (USB Video Class), which enables plug-and-play functionality across various operating systems.}
    \vspace{-4mm}
    \label{fig:architecture}
\end{figure}

\subsection{System Architecture}
\label{sec:architecture}
Our framework (\cref{fig:architecture}) emphasizes the preprocessing of multi-channel data prior to transmission via streaming protocols (e.g., RTSP). Unlike traditional methods, which rely on multiple channels to transmit multi-channel data, our method consolidates the data into a single channel. This is achieved by leveraging the capabilities of virtual 3D components within 3D engines (e.g., U3D), enabling efficient data fusion and transmission while maintaining low latency and high performance. The client directly receives the streamed data via the chosen protocol and plays the video. In our experiment, the RTSP protocol was selected for demonstration purposes. The framework is designed with a modular architecture, allowing developers to easily integrate other protocols such as RTCP, RTMP, WebRTC, HLS, etc., based on specific application requirements. It ensures that the chosen protocol on the streaming side decouples from our data preprocessing method in the framework, allowing the client-side to receive the streaming data based on the selected protocol. During the preprocessing of the multi-channel input data, each video stream is treated as a separate raw image, which is then mapped onto a virtual canvas within the 3D environment. The canvas is rendered in world space and captured by an in-world main camera to produce a single, consolidated video stream. This approach reduces the complexity associated with handling multiple video streams separately and minimizes redundant data transmission.

\subsection{Core Components}
Our approach involves constructing a theoretical model to establish the structure of data transmission, followed by video data preprocessing.

\textbf{Building the Model.} 
We formulate a model for low-latency live video data transmission, accounting for the relationship between external video capture devices and total streamed channels. Let \( F \) denote the number of live content channels and \( n \) the number of external video capture devices. Under USB-3.0 specifications, a single interface supports up to 256 devices \cite{usb3}, so \( n \in [0, 255] \).

Consider a main camera \( S \) within a virtual 3D scene. A virtual canvas \( c \) holds Raw Image display boards (\( \rho \)), each corresponding to a video stream from a single device, representing a non-interactive channel. The total non-interactive channels on a canvas are:
\vspace{-2mm}
\begin{align}
c(\rho) = \sum_{n=0}^{255} \rho_n
\end{align}

\vspace{-1mm}
Each interactive control group (\( \beta \)) on the canvas corresponds to an interactive board controlling stream display. 
Since each interactive board is paired with a Raw Image, the total channels on a single canvas are:
\vspace{-1mm}
\begin{align}
c = c(\rho) + c(\beta) = \sum_{n=0}^{255} \rho_n + \sum_{n=0}^{255} \beta_n = \sum_{n=0}^{255} (\rho_n + \beta_n)
\end{align}

\vspace{-1mm}
The scene includes two canvases: \( c_1 \) for video streams and \( c_2 \) for user interaction observation. The main camera \( S \) captures both, so \( S(c) = c_1 + c_2 \). The total channels \( F \) are:

\begin{align}
F(n) = S\left( c_1 \left( \sum_{n=0}^{255} \rho_n \right) + c_2 \left( \sum_{n=0}^{255} \beta_n \right) \right)
\end{align}

This model ensures multi-channel streaming achieves single-channel latency by consolidating streams before transmission. Equation (3) is derived by aggregating the channels from both canvases, captured by \( S \), which fuses video and interactive data into a single stream.

\textbf{Video Data Preprocessing.} 
The preprocessing component captures, renders, and encodes video streams within the 3D space. It creates Raw Image objects for each video channel, merges them onto a virtual canvas, and renders the canvas in world space, ensuring efficient multi-channel rendering and spatial transmission. Adaptive video encoding/decoding can be integrated, and functionalities like peer-to-peer communication, broadcasting, and text messaging are supported. \Cref{alg:streaming_setup} details the preprocessing and fusion process.

\vspace{-2mm}

\begin{algorithm}[H]
\caption{Streaming Side Setup}
\label{alg:streaming_setup}
\begin{algorithmic}[1]
    \Statex \textcolor{gray}{\textbf{Set up Recording Area:}}
    \State $\mathcal{S}_{3D} \gets \text{InitNewScene}()$ 
    \Comment{Init 3D scene}
    \State \text{SetRootGameObjects}$(\mathcal{S}_{3D})$ 
    \Comment{Set scene hierarchy}
    \State $\mathcal{C}_{ws} \gets \text{CreateCanvas}(\mathcal{W}\text{orldSpace}, \text{ScreenSize})$ 
    \Comment{Create canvas}
    \State $\mathcal{C}_{wsc} \gets \text{AddWorldSpaceCamera}(\mathcal{C}_{ws})$ 
    \Comment{Add canvas camera}
    \State $\mathcal{H}_{cameras} \gets \text{AddMultiCamHolder}(\mathcal{C}_{wsc})$ 
    \Comment{Add multiple cameras}
    \Statex \textcolor{gray}{\textbf{Set up Interaction Observer Module:}}
    \State $\mathcal{C}_{interaction} \gets \text{SetInteractionCanvas}(\text{RenderMode} \gets \mathcal{S}\text{creenSpaceOverlay}, \text{ScreenSize})$ 
    \Comment{Set interaction canvas}
    \State \text{Duplicate}$(\mathcal{H}_{cameras}.\text{Pages}) \rightarrow \mathcal{C}_{interaction}$ 
    \Comment{Copy camera pages}
    \Statex \textcolor{gray}{\textbf{Fuse All Cameras' Data into 1 Channel:}}
    \State $\mathcal{D}ata_{fused} \gets \text{WebcamVideoPlayer.Fuse}()$ 
    \Comment{Init data fusion}
    \Statex \quad \textbf{for} $i = 1$ \textbf{to} $n$ \textbf{do}
    \Statex \quad \quad $\mathcal{D}ata_{fused} \gets \mathcal{D}ata_{fused} + \mathcal{C}_{interaction}(i)$ 
    \Comment{Add interaction data}
    \Statex \quad \textbf{end for}
    \Statex \quad \textbf{for} $j = 1$ \textbf{to} $n$ \textbf{do}
    \Statex \quad \quad $\mathcal{D}ata_{fused} \gets \mathcal{D}ata_{fused} + \mathcal{H}_{cameras}(j)$ 
    \Comment{Add camera data}
    \Statex \quad \textbf{end for}
    \State \text{RTSP.Send}$(\mathcal{D}ata_{fused})$ \Comment{Send fused stream}
\end{algorithmic}
\end{algorithm}


Our framework leverages U3D to optimize multi-channel video data processing and transmission. By mapping streams onto virtual canvases and consolidating them via in-world cameras, it minimizes redundant processing and network data volume, enhancing latency reduction and scalability for applications like virtual events and VR/AR/MR content delivery.

\section{Evaluation}
To validate our framework, we conducted comprehensive evaluations focusing on four critical performance metrics: latency, synchronization accuracy, scalability, and user interaction responsiveness. We compared our approach to existing works to contextualize performance. Evaluations were conducted in a stable campus network environment with 73.65 Mbps upload and 54.06 Mbps download bandwidth, aligning with Wi-Fi 6 capabilities.

\subsection{Latency}
\label{sec:latency}
\begin{figure}[ht]
    \centering
    \includegraphics[width=0.45\textwidth]{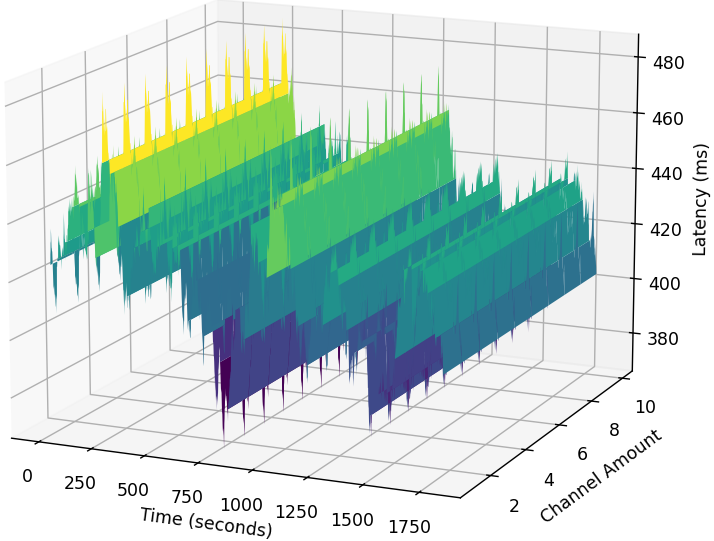}
    \vspace{-2mm}
    \caption{Latency measured over a 30-minute period with different numbers of input camera devices functioning as separate channels for live video streaming.}
    \label{fig:latency3d}
\end{figure}

Latency, the delay between video capture and playback (\( Latency = t_{render} - t_{capture} \)), is key for real-time applications \cite{uitto2021evaluation}. We varied input cameras from 1 to 10 (\cref{fig:latency3d}), showing consistent latency (~230 ms) regardless of channel count, confirming effective stream consolidation.

\vspace{-2mm}

We compared our approach with baselines: Sun et al. \cite{sun2021towards} (2314 ms), \textit{Learn2Adapt-LowLatency} \cite{karagkioules2020online} (1040 ms), and Zhang et al. \cite{zhang2021performance} (734 ms). Our method achieved a 68.7\% latency reduction over Zhang et al. for single-channel streaming (\cref{tab:latency}). Unlike baselines, where latency increases with channels, our framework maintained consistent performance.

\vspace{-2mm}

\begin{table}[h]
    \centering
    \caption{Latency Comparison with Baseline Schemes (ms).}
    \vspace{-2mm}
    \label{tab:latency}
    \begin{tabular}{@{}lccc@{}}
        \toprule
        \textbf{Channel Count} & \textbf{Single} & \textbf{Multi-Channel (5)} & \textbf{Multi-Channel (10)} \\
        \midrule
        Sun et al.\cite{sun2021towards}             & 2314            & 3462                       & 5645                       \\
        Learn2Adapt\cite{karagkioules2020online}    & 1040            & 2085                       & 3203                       \\
        Zhang et al.\cite{zhang2021performance}     & 734             & 1788                       & 3867                       \\
        Ours                                        & 230             & 231                        & 235                        \\
        \bottomrule
    \end{tabular}
\end{table}

These findings highlight the effectiveness of our framework’s design, which consolidates multiple video streams efficiently, minimizing delays and providing low-latency performance across a range of scenarios.

\subsection{Synchronization Accuracy}
\label{sec:sync_accuracy}
Synchronization accuracy measures the time difference between corresponding frames from multiple video streams. Let \(t_{channel_1}\) and \(t_{channel_2}\) denote the timestamps of frames from two different channels rendered on the client. The synchronization offset is calculated as \(\Delta t = |t_{channel_1} - t_{channel_2}|\). Since all channels are merged into a single stream, \(\Delta t\) is zero.

\subsection{Scalability}
Scalability is evaluated by measuring the system's performance as the number of video streams increases. Let \(n\) represent the number of streams, namely the number of input devices, and \(L(n)\) be the latency as a function of \(n\). The system's scalability is considered strong if \(L(n)\) stays almost stable within a tolerable range with increasing \(n\). The computational load on the system (e.g., CPU/GPU consumption) is denoted by \(U_{CPU}(n)\) and \(U_{GPU}(n)\), and we expect these to increase linearly: 
\(U_{CPU}(n), U_{GPU}(n) \propto n\). 

\begin{figure}[h]
    \centering
    \vspace{-2mm}
    \includegraphics[width=\linewidth]{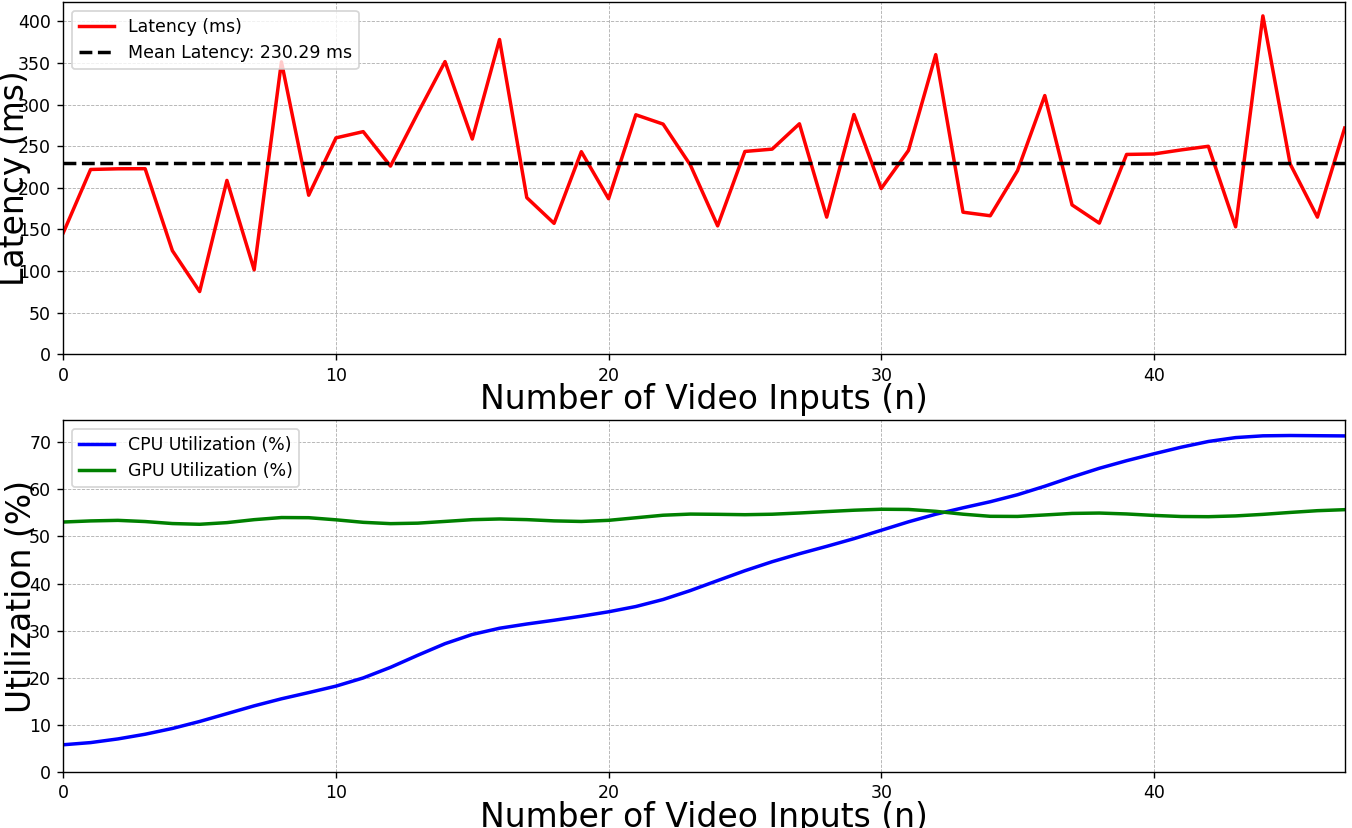}
    \vspace{-6mm}
    \caption{Evaluation of system performance as the number of input devices increases to 50. The average latency is 230.29 ms. GPU consumption remains stable with minimal fluctuations, while CPU consumption rises progressively with more input devices. The GPU consumption stays almost still because the online-rendering is conducted if only the video is displayed. This indicates efficient GPU resource management and increasing CPU demands. Testing was conducted on machines with an NVIDIA GeForce GTX 1050 GPU and an Intel Core i7-7700HQ CPU.}
    \label{fig:scalibility}
\end{figure}

Our evaluation (\cref{fig:scalibility}) on latency demonstrates that our method (230.29ms) surpasses \cite{karagkioules2020online} (1040ms) in single-channel live video streaming significantly, while also outperforming some traditional methods (\cite{shuai2014low, swaminathan2011low, wei2014low, gul2020cloud}), as well as more recent methods (\cite{wang2021multilive, lee2020groot, kim2020neural}). Notably, as the number of input devices increases, our approach maintains consistent bandwidth, showcasing its scalability and robustness. This is due to its design, which merges multiple channels into one before the streaming process.

\subsection{User Interaction Responsiveness}
User interaction responsiveness is measured by the time delay between user input and the corresponding action in the video stream. Let \(t_{input}\) be the time of user input, and \(t_{action}\) be the time when the action is reflected in the live stream. The responsiveness is then defined as: \(Response Time = t_{action} - t_{input}\). 
Lower response times indicate better interaction responsiveness.

\begin{figure}[ht]
    \centering
    \includegraphics[width=\linewidth]{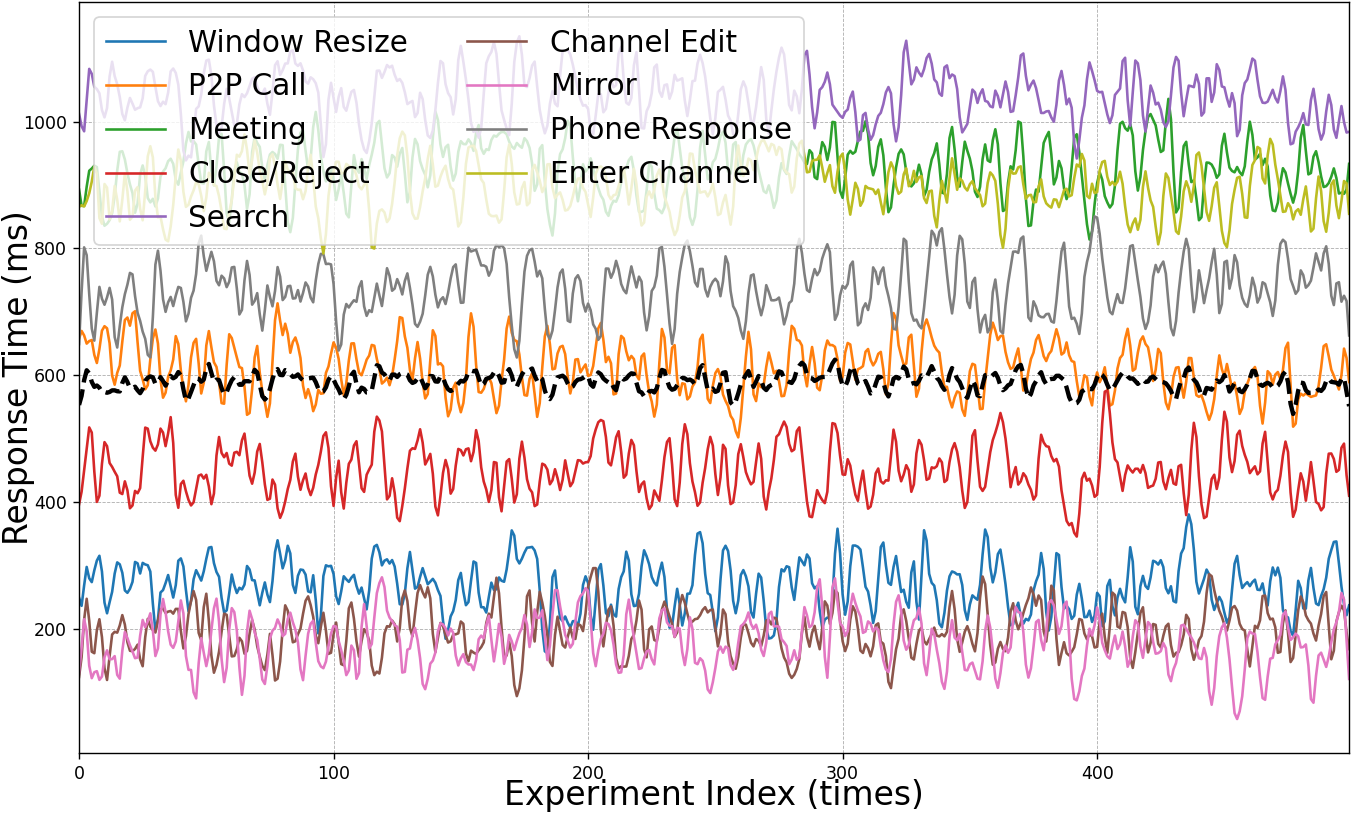}
    \vspace{-6mm}
    \caption{Nine modules were selected for evaluation, with their response times predominantly ranging from 50 ms to 1300 ms. The mean response time, represented by a dashed line, is approximately 600 ms. Given the significant fluctuations observed in the recorded data, Gaussian smoothing \cite{ito2000gaussian} was applied to reduce noise and variability, utilizing a parameter (\(\sigma\)) of 1 for visualization purposes.
    }
    \label{fig:interaction_smoothed}
\end{figure}

In the experiment, we performed over 500 interactions across 9 modules of our live streaming system implemented in U3D and recorded the corresponding interaction data (\cref{fig:interaction_smoothed}).

These results demonstrate our framework’s responsiveness and robustness across diverse conditions. Our method achieves efficient response times, ensuring a smooth user experience. Its computational efficiency, indicated by stable GPU usage and linear CPU growth, validates its scalability and resource optimization for modern applications.

\section{Limitation}  
Despite the notable advantages demonstrated by our multi-channel live streaming framework, certain limitations warrant consideration. A key trade-off observed lies in the reduction of video quality under multi-channel conditions. While the framework efficiently fuses data streams to optimize transmission bandwidth, this design inherently results in a loss of quality as the number of input channels increases. Specifically, the consolidation process maintains fixed video sizes for transmission, leading to a proportional decline in visual fidelity with an increasing number of channels when end-users simultaneously request the display of multiple channels.

This limitation reflects the inherent compromise between maintaining low latency and achieving optimal video quality, particularly in scenarios involving a large number of channels associated with requests for multiple displays at the same time. While our framework achieves superior performance metrics in latency, synchronization, scalability, and user interaction responsiveness, the reduction in video quality may pose challenges in use cases where high-definition output is critical, such as medical imaging or high-resolution broadcasting.

Future work should explore adaptive encoding techniques or advanced compression algorithms tailored for multi-channel environments to mitigate this issue. Additionally, implementing quality-preservation strategies, such as dynamic resource allocation or channel prioritization based on content importance, may enhance the framework’s overall effectiveness and broaden its applicability in quality-sensitive domains.

\section{Conclusion} 
Our 3D framework for enhancing low-latency multi-channel live video streaming, which integrates video within a virtual 3D space, demonstrates advantages over traditional live streaming methods across several key performance metrics. This framework achieves considerable reductions in latency, high synchronization accuracy, and robust scalability. These findings confirm the effectiveness of incorporating multi-channel video streams within a 3D virtual environment and underscore the potential of this approach for various multi-channel live video streaming applications, particularly in contexts that demand low-latency video delivery.

\section*{Acknowledgement}
This work was supported by the Ministry of Education of the People's Republic of China under the Industry-University Cooperative Education Program (BJUT Contract No. 40059000202520) through the New Engineering Construction Project Cooperation Agreement.

\bibliographystyle{IEEEbib}
\bibliography{icme2025references}

\end{document}